# CobWeb — a toolbox for automatic tomographic image analysis based on machine learning techniques: application and examples

**Swarup Chauhan[1,2*], Kathleen Sell[2,5], Frieder Enzmann[2], Wolfram Rühaak[3], Thorsten Wille[4], Ingo Sass[1], Michael Kersten[2]**

[1]*Institute of Applied Geosciences, University of Technology, Darmstadt 64287, Germany*

[2]*Institute for Geosciences, Johannes Gutenberg-University, Mainz 55099, Germany*

[3]*Federal Institute for Geosciences and Natural Resources (BGR), Hannover 30655, Germany*

[4]*APS Antriebs-, Prüf- und Steuertechnik GmbH, Götzenbreite 12, Göttingen-Rosdorf 37124, Germany*

[5]*igem – Institute for Geothermal Ressource Management, Berlinstr. 107, Bingen 55411, Germany*

**Abstract.** In this study, we introduce CobWeb 1.0 which is a graphical user interface tailored explicitly for accurate image segmentation and representative elementary volume analysis of digital rock images derived from high resolution tomography. The CobWeb code is a work package deployed as a series of windows executable binaries which use image processing and machine learning libraries of MATLAB®. The user-friendly interface enables image segmentation and cross-validation employing K-means, Fuzzy C-means, least square support vector machine, and ensemble classification (bragging and boosting) segmentation techniques. A quick region of interest analysis including relative porosity trends, pore size distribution, and volume fraction of different phases can be performed on different geomaterials. Data can be exported to ParaView, DSI Studio (.fib), Microsoft® Excel and MATLAB® for further visualisation and statistical analysis. The efficiency of the new tool was verified using gas hydrate-bearing sediment samples and Berea sandstone, both from synchrotron tomography datasets, as well as Grosmont carbonate rock X-ray micro-tomographic dataset. Despite its high sub-micrometer resolution, the gas hydrate dataset was suffering from edge enhancement artefacts. These artefacts were primarily normalized by the dual filtering approach using both non-local means and anisotropic diffusion filtering. The desired automatic segmentation of the phases (brine, sand, and gas hydrate) was thus successfully achieved using the dual clustering approach

# 1 Introduction

Despite the availability of both commercial and open source software for digital rock physics (DRP) analysis as compiled in Figure 1, an ideal tool for accurate automatic image analysis at ambient computational performance is difficult to pinpoint. The best practice so far among researchers is to alternate between different available software tools and to synthesize the different datasets using home-brew workflows. Porosity and particularly permeability can vary dramatically with small changes in segmentation, as significant features can be lost when thresholding greyscale tomography images to binary images, even if using the most advanced data acquiring techniques like synchrotron tomography (1). Our new CobWeb 1.0 visualisation and image analysis toolkit addresses some of the challenges of selecting representative elementary volume (REV) for X-ray computed tomography (XCT) datasets reported earlier (1-7). It is customized to perform image analysis and accurate greyscale phase segmentation of reconstructed high resolution XCT and synchrotron tomographic datasets. As the only one currently available, it is based on machine learning techniques of excellent performance for segmentation analysis as detailed previously (8, 9). This software tool package was developed on a MATLAB® workbench and can be used as Microsoft Windows standalone executable (.exe) files or as a MATLAB® plugin. In this paper, we demonstrate exemplarily 3D tomographic REV analysis of Berea sandstone, Grosmont carbonate rock, and gas hydrate-bearing sediment datasets. For the latter geomaterial, Sell et al. (10) highlighted problems with the edge enhancement (ED) effect and recommended image morphological strategies to compensate for this artefact. In this paper, we suggest a strategy to eliminate ED artefacts using the same dataset with the machine learning approach. The respective MATLAB code is provided in the appendix. The salient features of CobWeb 1.0 and its overall framework are described in Section 2. Section 3 highlights the image segmentation techniques used in Section 4 for REV, relative porosities, and pore size distribution analysis.

# 2 CobWeb 1.0

## 2.1 Salient Features

The word *Cobweb* means "a tangled three-dimensional spider web", i.e. something resembling a cobweb in delicacy or intricacy (Oxford Dictionaries). According to Marriam-Webster, it may also mean something that entangles obscures or confuses, as is the philosophy of machine learning - elegant, sophisticated yet stochastic and confusing. This inspired us to name our software tool CobWeb. The first version enables reading and processing of (reconstructed) XCT files in both tiff and raw formats. The graphical user interface (GUI) is embedded with visual inspection tools for zooming in/out, cropping, color, and scale, and assisting in the visualisation and interpretation of 2D and 3D stack data. Noise filters such as non-local means, anisotropic diffusion, median and contrast adjustments are implemented to increase signal-to-noise ratio. The user can choose from a series of five different segmentation algorithms, namely K-means, Fuzzy C-means (unsupervised), least square support vector machine (supervised), bragging and boosting (enable classifiers) for accurate automatic segmentation and cross-validation. Relevant material properties like relative porosities, pore size distribution, volume fraction (pore, matrix, mineral phases) can be quantified and visualized as graphics output. The data can be exported into different file formats such as Microsoft® Excel (.xlsx), MATLAB® (.mat), ParaView (.vkt) and DSI studio (.fib). The current version is supported for Micosoft® Windows PC operating systems (Windows 7 and 10).

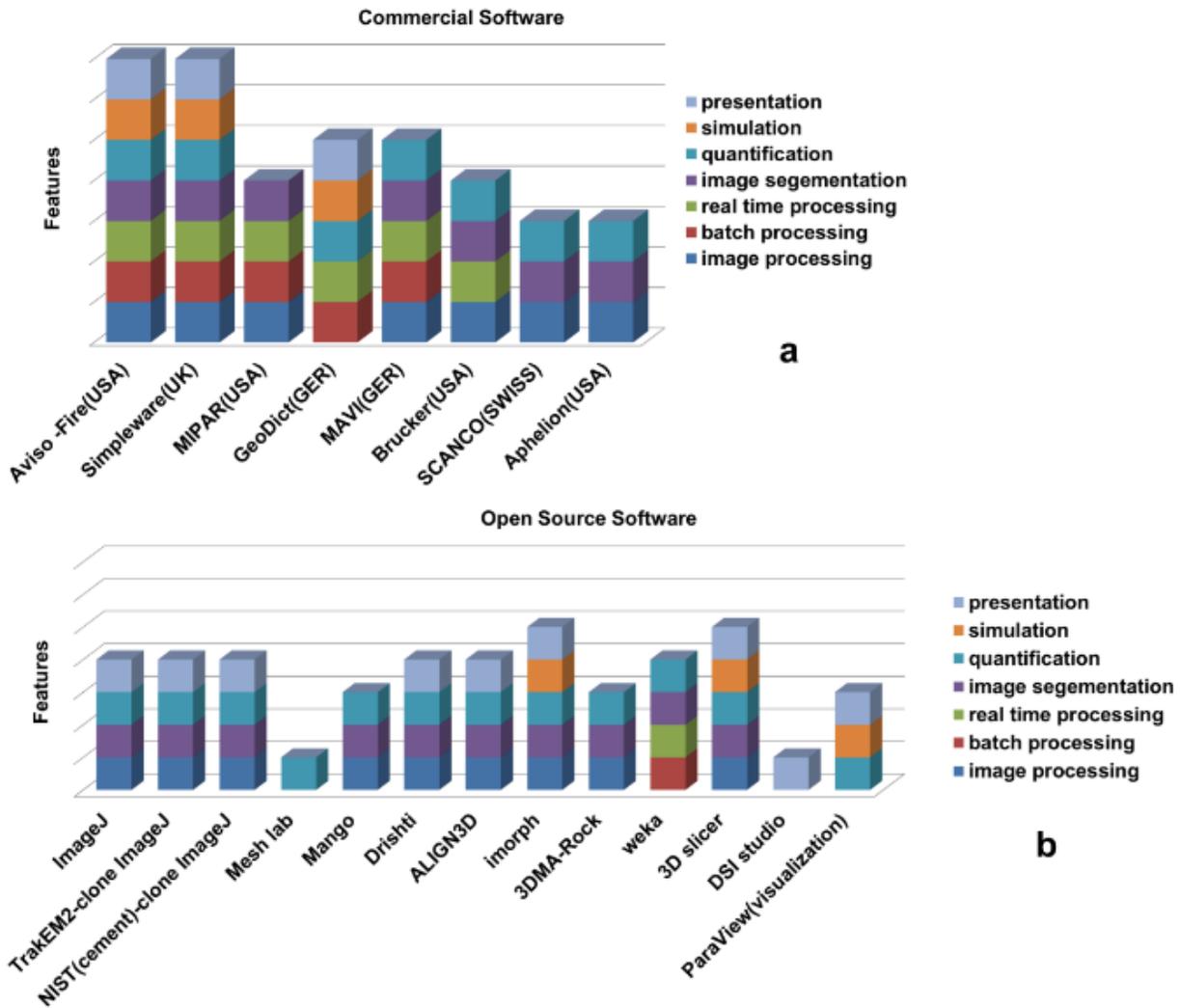

**Figure 1.** Market survey of the currently available commercial software (a) and open source software (b) assisting in digital rock physics analysis with features as indicated in the legend.

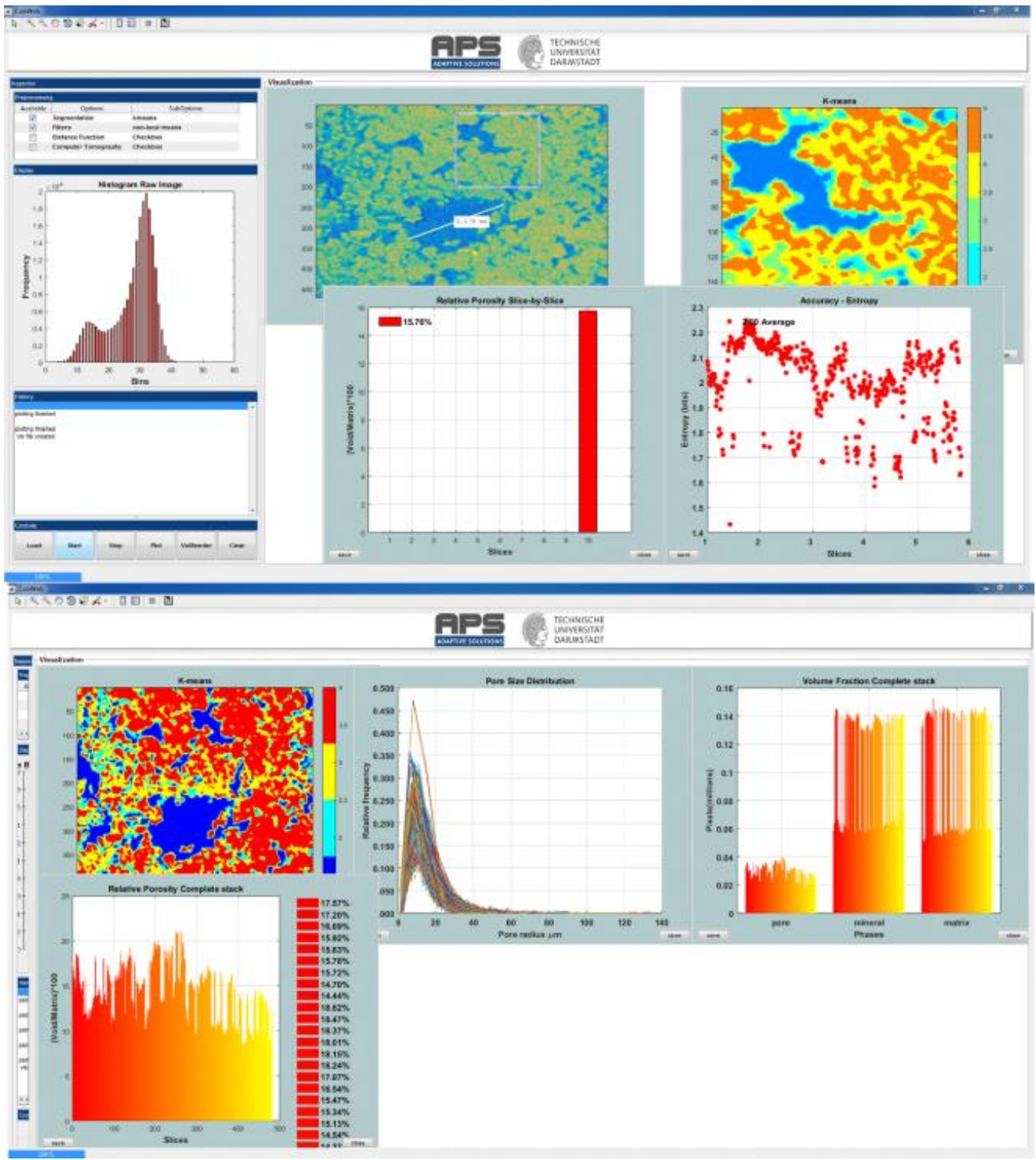

**Figure 2.** Screenshots of the CobWeb GUI. XCT stack of Grosmont carbonate rock is shown as an example of representative elementary volume analysis. The top panel displays the XCT raw sample, the K-means segmented ROI, and the porosity of single slice No. 10. The bottom plot shows pore size distribution of the complete REV stack, the relative porosity and volume fraction, respectively.

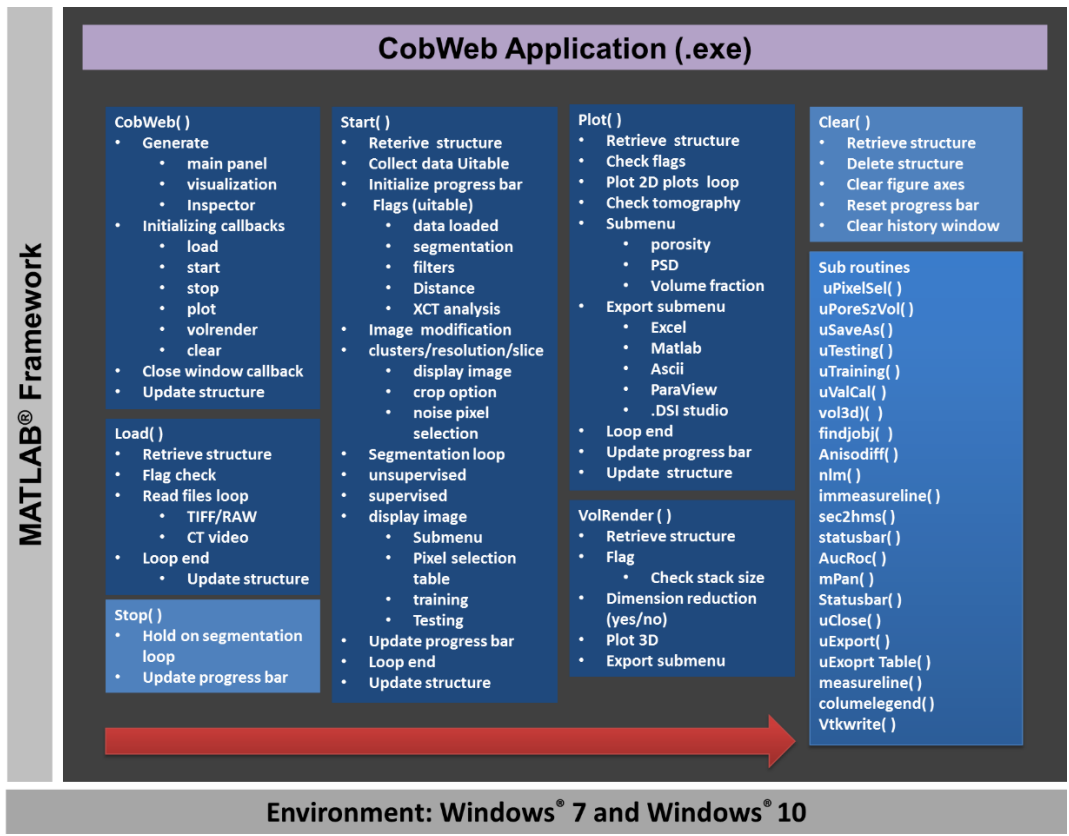

**Figure 3.** The general workflow of the CobWeb software tool, where the arrow denotes the series in which different modules are compiled and executed. A separate file script is used to generate .dll binaries and executables.

## 2.2 Window Panels

The main GUI window panel divides into three main parts (Figure 2), the tool menu strip, the inspector panel, and the visualisation panel. The tool strip contains menus for zoom in and out, pan, rotate, point selection, color bar, legend bar, and measurement scale functionalities. The inspector panel is divided into subpanels where the user can configure the initial process settings such as segmentation schemes (supervised, unsupervised, ensemble classifiers), filters (contrast, non-local means, anisotropic filter, *fspecial*), and distance functions (link distance, Manhattan distance, box distance) to assist segmentation and geometrical parameter selection for image analysis (REV, porosity, PSD, volume fraction). The display subpanel records and displays the 2D video of the XCT stack and the histogram of the raw image(s). The history subpanel is a *uilistbox* that displays errors, processing time/status, processing instruction, files generated/exported and executed callbacks. The control subpanel is an assemblage of *uibuttons* to initialize the XCT data analysis process and the progress bar. The visualisation panel is where the results are displayed in several resized windows, which can be moved, saved and deleted. The window panels displayed in the visualisation module are embedded with *uimenu* and *submenu* to export, plot and calculate different variables like porosity, PSD, volume fraction, entropy, or receiver operational characteristics. To get the desired user functionalities, MATLAB® internal *uilibaries* were used and where inadequate. Therefore, numerous specific adaptions were adopted from Yair Altman's undocumented Matlab website and the Matlab File Exchange community. Specifically, the GUI Layout Toolbox from David Sampson has been used to configure the CobWeb GUI layout; the *uitable*, which uses the MATLAB java-component was designed using *uitable* customisation *reoprt* provided by (10).

## 2.3 Overall Framework

An overview of the different modules of the CobWeb toolkit are compiled in Figure 3, and the arrow displayed indicates the series in which they are executed. The advantage of using MATLAB® is the access to the structure and respective variables, which are used for further investigations. As a stand-alone, the GUI can be executed on different PC and HPC clusters without any license issues. The CobWeb 1.0 framework can be broadly divided into three modules

### 2.3.1 Control Module

In the control module, the CobWeb menu creates the main figure panel, assembles the size/position of the panel and subpanel windows, initializes the control buttons and generates a main structure. Ideally, any button can be activated after the GUI is displayed, but an exception will be displayed in the history window, highlighting the next step. That step is to load data where the *Load* function checks the file properties, loads the data in tiff and raw formats, creates and displays 2D video of the selected stack, saves the video file in the current folder, and updates the respective variables to the main structure. The *Stop* function ends the execution. However, when the processing is inside a loop, the *Stop* function breaks the loop after the i-th iteration.

### 2.3.2 Analysis Module

The *Start* function is in the analysis category and is a densely nested function, where the bullet points and the indented bullet points represent the outer and the inner nested loops, respectively (Figure 3). Initially, data is gathered and a sanity check is performed to evaluate whether all the options are correctly selected. If the conditions are not satisfyied, an exception alert pops up in the *History* panel, highlighting the error and offering an alternative process. The second loop is the image modification loop, where initially the user input is required. This input comprises a desired cluster number, the given image resolution, and the slice number. Afterwards, the chosen slice is displayed in 2D format, and in a separated window. Following this, an option of ROI selection is proposed to be accepted or denied. If the user accepts, the REV will be cropped and updated to the main structure and the cropped ROI is displayed in a separate window as a slice.

Based on the option selected in the *uitable*, the respective unsupervised and supervised loop is initialized. If LSSVM or Bragging and Boasting is chosen as the segmentation scheme, a right click *uimenu* is initialized with the options of pixel selection, training, and testing. On pressing the *Pixel Section* option, the subroutine *uPixelSel()* initiates a *uitable* window representing the columns *Clusters*, *Features*, *X-Coordinate*, and *Y-Coordinate*, and requiring user inputs in the respective columns. The user can explore interesting features (pores, minerals, matrix, noise/specks) in the 2D image windows and collect the data using zoom in, zoom out, and data cursor tools. Once the respective features' X,Y coordinates are fed in the *uitable*, the data has to be arranged and exported for training and testing. This is fulfilled by pressing the export button which initiates the subroutine *uExportTable()*. The export subroutine collects a total of 36 (6 x 6) pixel values in the perimeter of the X,Y coordinates of the respective features given in the *uitable*. Thereafter, with the training and testing options, respective models (LSSVM, Ensemble Classifiers) are trained using the pixel values of the representative slice. The classification is then performed on the 3D stack, and the main structure is updated. In the case of unsupervised techniques based on the option selected for image filtering, segmentation, and distance function, the complete stack is processed. For FCM, the user is given an option to choose the membership criteria (7) between the range of one to two (decimal values).

The progress can be monitored in the progress bar, the color of the control buttons (red to grey) and the *History* window, which gives the related information on processing time, segmentation scheme, and filter options executed.

### 2.3.3 Visualisation Module

The visualisation module consists of the plot function and volume rendering function. The plot function is a densely nested function, where the bullet points and the indented bullet points in represent the outer and the inner nested loops, respectively. The execution of the *Plot()* callback accesses the main structure, identifies the representative slice number, and plots the segmented 2D image in a new window. The displayed window is embedded with right click menu and submenus, namely:

- Porosity → Porosity, Pore Size Distribution
- Validate → Entropy, Receiver Operation Characteristics, 10-fold Cross Validation
- Export Stack → Paraview Format

When choosing the desired options mentioned above, their respective subroutines (*uPoreSzVol*, *uCalVal*, *uExport*) are started and the desired results are plotted as shown in Figure 2. The user has the option to choose a single slice or the complete stack. The visualized parameter distributions display in a separate window and can further be exported using right-click menu into Excel, ASCII, or MATLAB formats.

The *Export Stacks* option bundles the complete stack in ParaView code format (.vkt files) using the subroutine function *Vtkwrite*. When the VolRender button is activated, the subroutine checks for the size of the stack and offers an option to perform volume reduction of the greyscale values. If ignored, depending on the computer RAM capacity, it may take a relatively long time to plot a high quality volume rendered figure of the 3D stack. If accepted, the pixel information is reduced 10-fold, which fastens the plotting process, but the image quality is hampered. Therefore, CobWeb 1.0 offers the option to export the stack and visualize it using ParaView or DSI studio.

### 2.3.4 Miscellaneous Options

CobWeb 1.0 has a clear button to reset the CobWeb by delegating back to the main panel. This corresponding function (Clear) clears the figure axis, the history window panel, and resets the progress bar. The plot windows embed, save, and close *fbuttons* which can save current plots to different image formats (bmp, matlab, tiff, Jpeg, ect.) and closes the current window.

## 3 Tomography Datasets Used for Evaluation of CobWeb 1.0

### 3.1 Gas Hydrate-Bearing Sediments

The in-situ synchrotron-based tomography experiment and post-processing of synchrotron data conducted to resolve the microstructure of gas hydrate-bearing (GH) sediments is given in detail by (11-12), and (9). In brief, the tomographic scans were acquired with a monochromatic X-ray beam energy of 21.9 KeV at Swiss Light Source synchrotron facility (Paul Scherrer Institute Villigen, Switzerland) using the TOMCAT beamline (Tomographic Microscope and Coherent Radiology Experiment) (13). Each tomogram was reconstructed from sinograms by using the gridded Fourier transformation algorithm (14). Adjacent to this, a three-dimensional stack was derived resulting in an image size of 2560 x 2560 x 2160 voxels with a voxel resolution of 0.74 μm and 0.38 μm at 10-fold and 20-fold optical magnification, respectively.

### 3.2 Grosmont Carbonate Rock

The datasets of the Grosmont carbonate rock were obtained from the *GitHub* FTP server (http://github.com/cageo/Krzikalla-2012) provided for the benchmark project reported by Andrä et al. (15, 16). They acquired the Grosmont carbonate rock for their benchmark test from the Grosmont Formation in Alberta, Canada. The Grosmont Formation was deposited during the Late Devonian and is divided into four facies members, LG UG-1, UG-2, and UG-3 from the bottom up. The sample was taken from UG-2 facies and is mostly composed of dolomite and karst breccia (17, 18). Laboratory measurements of porosity and permeability reported in Andrä et al. (16) are 0.21 (21 %) and 150 mD ─ 470 mD, respectively. The Grosmont carbonate dataset was measured at the high-resolution X-ray computer tomographic facility of the University of Texas with an Xradia MicroXCT-400 instrument (ZEISS, Jena, Germany). The measurement was performed using 4x objective lenses, 70 kV polychromatic X-ray beam energy and a 25 mm CCD detector. The tomographic images were reconstructed from the sinograms using proprietary software and corrected for the beam hardening effect as typical for lab-based polychromatic cone-beam X-ray instruments (18). The retrieved image volume was cropped to a dimension of 1024 x 1024 x 1024 with a voxel resolution of 2.02 μm.

*3.3 Berea Sandstone Rock*

The Berea sandstone dataset was also obtained from the *GitHub* FTP server provided for the benchmark project reported by Andrä et al. (15, 16). The Berea sandstone sample plug was acquired from Berea sandstone TM Petroleum Cores (Ohio USA). Porosity values of around $\varphi = 0.20$ (20 %) were obtained using a Helium pycnometer 1330 (Micromeritics Instrument Corp., Germany) and a Pascal 140+1440 Mercury porosimeter (Thermo Electron Corporation, Germany) as described by Giesche (20). The permeability reported in the same benchmark test (16) ranges between 200 mD and 500 mD. Machel and Hunter (17) reported for this sample a mineral composition of Ankerite, Zircon, K-feldspar, Quartz, and Clay using a polarized optical microscope and a scanning electron microscope. The synchrotron tomographic scans of Berea sandstone were also acquired at the SLS TOMCAT beamline (13). The beam energy was monochromatized to 26 keV for optimal contrast with an exposure time of 500 ms. This resulted in a 3D tomographic stack of dimension 1024 x 1024 x 1024 and voxel resolution of 0.74 μm.

## 4 Results and Discussion

*4.1 Image Pre-Processing*

XCT and synchrotron tomographic datasets were used to validate the functionality of CobWeb 1.0. A total of 12 region of interests (ROIs) were thus investigated to determine suitable REVs. Figure 4 shows the ROIs selected for the Berea, Grosmont and gas hydrate samples. Image pre-processing is one of the essential and precautionary steps before image segmentation (21, 22). Image filtering techniques such as blur, background intensity variation and contrast help in reducing artefacts. Image denoising filter such as median filter, non-local means filter, and anisotropic diffusion filter can assist in lowering the phase misclassification and improving the convergence rate of automatic segmentation schemes. In the case of XCT volume stack of Berea sandstone, the 3D reconstructed raw images ($1024^3$) had sufficiently high resolution and contrast, thus they did not show any noticeable change on using the above mentioned filters. However, with XCT images ($1024^3$) of the Grosmont carbonate rock, non-local means filtering yielded better visualisation and performance results compared to anisotropic diffusion filter.

*4.1.1 Dual Filtering of Gas Hydrate-Bearing Sediments*

Due to ED artefacts affecting the quality of the hydrate-bearing sediment tomograms, the data had to undergo a data post-processing routine to enhance the image quality. Details of the image enhancement are published in Sell et al. (9). In brief, several image enhancement techniques were tested in preliminary studies including filters and filter combinations to gain best-fit results for further numerical simulations. To our knowledge, the combination of the non-local means and the anisotropic diffusion filter, both implemented in Avizo (ThermoFisher Scientific), works best for the given GH data.

The concept of the anisotropic diffusion (AD) filter is to smooth out noise in predefined areas of an image, but also to stop at sharp edges representing boundaries between phases. This way, edges and sharp boundaries between phases are preserved, and image noise is significantly reduced (23, 24). A comparison of the current voxel with its six neighbors takes place, and diffusion is fulfilled when the threshold stop criterion is not exceeded. If the difference between one voxel and its six adjacent neighbors exceeds the given value no diffusion takes place. Another option to control the diffusion process of the filter is to reduce or increase the diffusion time. The parameter number of iterations defines how often the algorithm will be used on the data. The bigger this number, the more blurred the resulting image. Smoothing is performed by applying a Gaussian filter. For our investigations, the threshold stop criterion was set to the value 22,968 as this is the approximated transition of the grain phase to hydrate. AD was run on a CPU device with five iterations.

The non-local means filter (NLM) is a windowed version of the non-local means algorithm (25, 26). The main aim is to de-noise data based on comparing voxels for similarities in a selected window in which a new weight for a voxel is assigned. After a Gauss kernel was run on the weighted values, the new value will be assigned replacing the former grey values. The filter is most efficient if the image is affected by white noise. In Avizo the parameter window size, the local neighborhood, and the similarity value can be customized. Furthermore, the NLM filter is also an appropriate tool for salt-and-pepper de-noising caused by image sensor defects (27). For this study, the NLM filter was run in 3D mode on a CPU device. The *search window* was set to 21 and the *local neighborhood* to 6 at a similarity value of 0.71.

*4.2 Phase Segmentation*

*4.2.1 Grosmont Carbonate and Berea Sandstone Rock Samples*

The K-means algorithm was used for the segmentation of REV analysis of Berea and Grosmont rocks. K-means is one of the simplest unsupervised machine learning (ML) algorithms commonly used to address clustering problems (28, 29, and 7). The K-means algorithm iteratively calculates the Euclidean distance between the data points (pixel value) to its nearest centroid (cluster). The algorithm converges when the objective function, i.e. the mean square root error of Euclidean distance, reaches the minimum. This is when no further pixel is left to be assigned to the nearest centroid (cluster). However, the K-means algorithm has the tendency to terminate without identifying the global minimum of the objective function. Therefore, running the algorithm several times is recommended to increase the likelihood that the global minimum of the objective function will be identified. The performance of the K-means algorithm is strongly governed by the initial choice of the cluster centres (7).

The supervised ML techniques rely on features also termed as feature vectors (FVs). The FVs are sets of instances that represent descriptive information on which ML algorithm is used to train the classification model. They further identify these features in an unknown dataset and group them into respective classes. Least square support vector machine (LSSVM) is one such supervised ML technique, which in recent years has emerged as a reliable technique to segment digital rocks images (7). Khan et al. (30) provides concise description and MATLAB® code snippet for the implementation of the LSSVM library on XCT images, whereas Chauhan et al. (8) validated its best performance and accuracy in comparison to other common ML techniques. In practice, an FV is a group containing subsets of different pixel values. For example, the FV of class four is a group encapsulating pixel values corresponding to the pore, matrix rock, and noise. The pixel values were selected from a single 2D slice representative of the REV. This FV was used for training the classification model. The training performance was monitored using a 10 K-fold cross-validation technique (31-33).

*4.2.2 Gas Hydrate-Bearing Sediments ─ Dual Clustering*

The edge enhancement (ED) effect was significant in all reconstructed slices of the GH dataset. The ED effect was seen around the quartz grains mostly, with high and low pixel intensities adjacent to each other. The high intensity pixel (EDH) values were very close to GH pixel values, while the low intensity pixel (EDL) values showed a variance between noise and brine phase pixel values. Therefore, immediate segmentation performed on the pre-filtered GH datasets using CobWeb 1.0 resulted in misclassification. Further parameterising and tuning the unsupervised (K-means) and supervised (LSSVM) modules of CobWeb 1.0 specifically, distance function (i.e., functions euclidean distance *sqeuclidean*, sum of absolute differences *cityblock*, and *mandist*) and different permutation and combination between kernel type, bandwidth and cross-validation parameters, showed significant improvement, but the segmentation was still not optimal. The aim was to eliminate the ED features completely without altering the phase distribution between GH and the brine. This prompted development of a GH-specific workflow as explained below. The appendix provides the MATLAB® script for this workflow comprised of 6 steps:

- **Step 1: Filtering and REV selection**
  Four REVs of size $4 \times 700^3$ were cropped from the raw (16 bit) data stack. These REVs were dual-filtered using NLM and ASD filters (see section 4.1.1). Figure 6 shows the pre-filtered raw dataset.
- **Step 2: K-means clustering**
  In this step, K-means segmentation is performed on the REVs to label the phases into different classes. The class sizes were varied between three to twenty, and it was thus established that class seven captured all the desired phases (noise, edge enhancement low intensities (EDL), brine, quartz, edge enhancement high intensities (EDH), GH).
- **Step 3: Indexing**
  The pixel indices corresponding to the respective classes (desired phases) were extracted from the segmented slice(s). Thereafter, using these pixel indices as reference, corresponding pixel values were extracted from the 16-bit raw images. The obtained pixel values represent noise, EDL, brine, quartz, EDH, and GH phases in the raw images. Then, histogram distribution of the pixel values in each phase was plotted. The skewness of the histogram plots was inspected visually and mean and standard deviation for each histogram were calculated. If an overlap of pixel intensities was found in the different histograms (phases), step four was repeated.
- **Step 5: Rescaling raw REV**
  The pixel values corresponding to the phases, i.e brine, quartz, and GH, were replaced by their mean values, with an exception for EDH pixel values. The latter were replaced with the mean value of quartz. These assignments lead to optimal segregation of the phase boundaries in the raw dataset and finally to the elimination of the ED effect.
- **Step 6: K-means clustering**
  K-means segmentation with three classes was performed on the rescaled dataset to obtain the final result.

### 4.3 Representative Elementary Volume Analysis

The representative elementary volume (REV) can be defined as the smallest volume, which should ideally represent the average effective macroscopic behavior of the geomaterial. As a result, the transport of the effective parameters (mass, momentum, energy) mathematically modelled within the REV becomes independent of the sample size (34).

Figure 5 schematically explains the relationship between porosity and the volume of the porous media. In a small REV (region I), high fluctuation in porosity is caused by heterogeneity at the pore scale. As volume increases (region II), porosity starts to normalize above some $V_{min}$ value within a small standard deviation around a constant value of porosity. The porosity measured in this region is scale-independent, and an accurate representation of a large-scale system. The increase in REV value above a $V_{max}$ may result in increase/decrease in porosity related to increases in heterogeneity, associated with `macroscopic' volume features (region III) (34). For heterogeneous porous media, porosity theoretically lies in between region I and region III depending on the effective parameter under investigation; however, the determination of ideal region II for a real heterogeneous system may be difficult and subjective (2-6).

In particular while performing permeability tensor simulation using XCT data, the size of the minimum REV should be assessed based not only on porosity but also on geometrical parameters such as pore size distribution, void ratio, and coordinate number (5, 6). For this study, we investigated different ROIs and REV sizes between $300^3$ to $500^3$ resolution, and established that an REV of size 471 x 478 x 480 suited best. Figure 7 shows the REVs of Berea sandstone and Grosmont carbonate rock and their respective geometrical parameters.

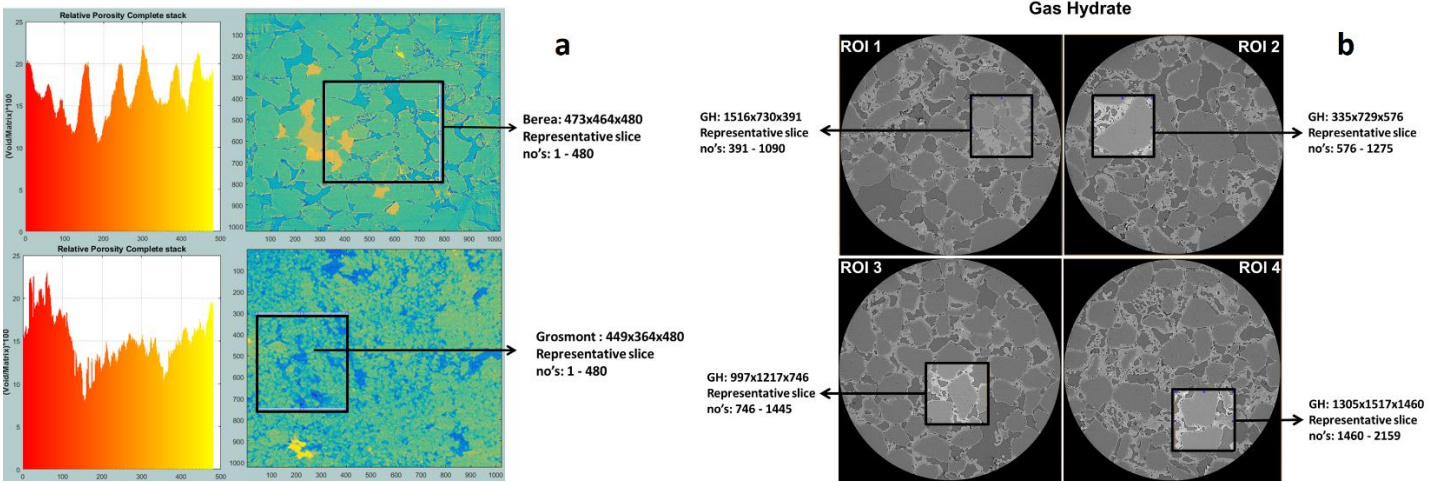

**Figure 4.** The most suitable REVs of Berea sandstone and Grosmont carbonate rock shown in panel and gas hydrate-bearing sediment datasets shown in the panel b.

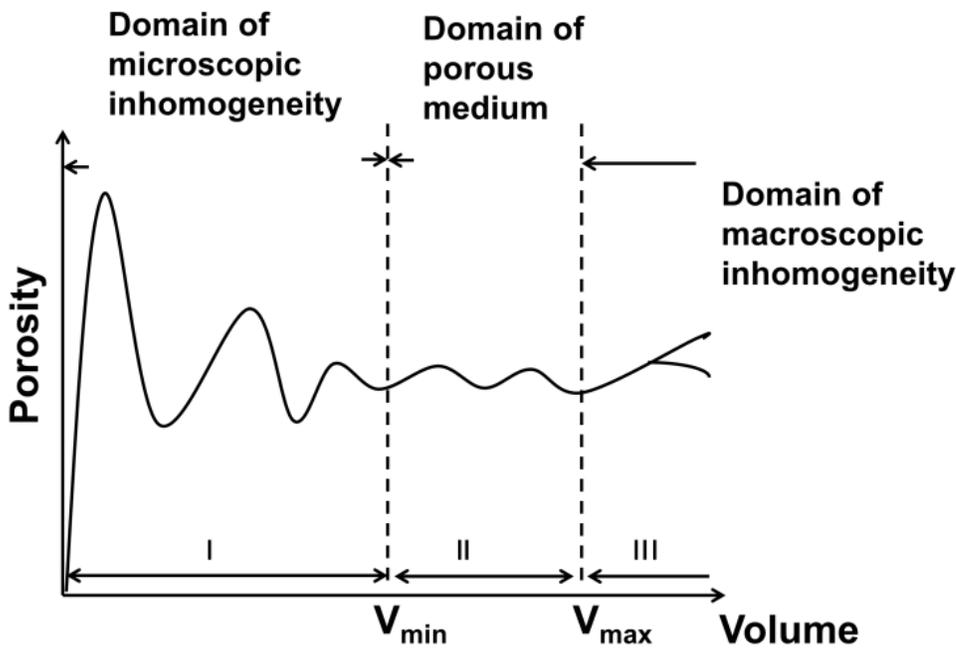

**Figure 5.** Schematic representation of the relationship between porosity (φ) and volume (V) of porous media. Bachmat and Bear (1986).

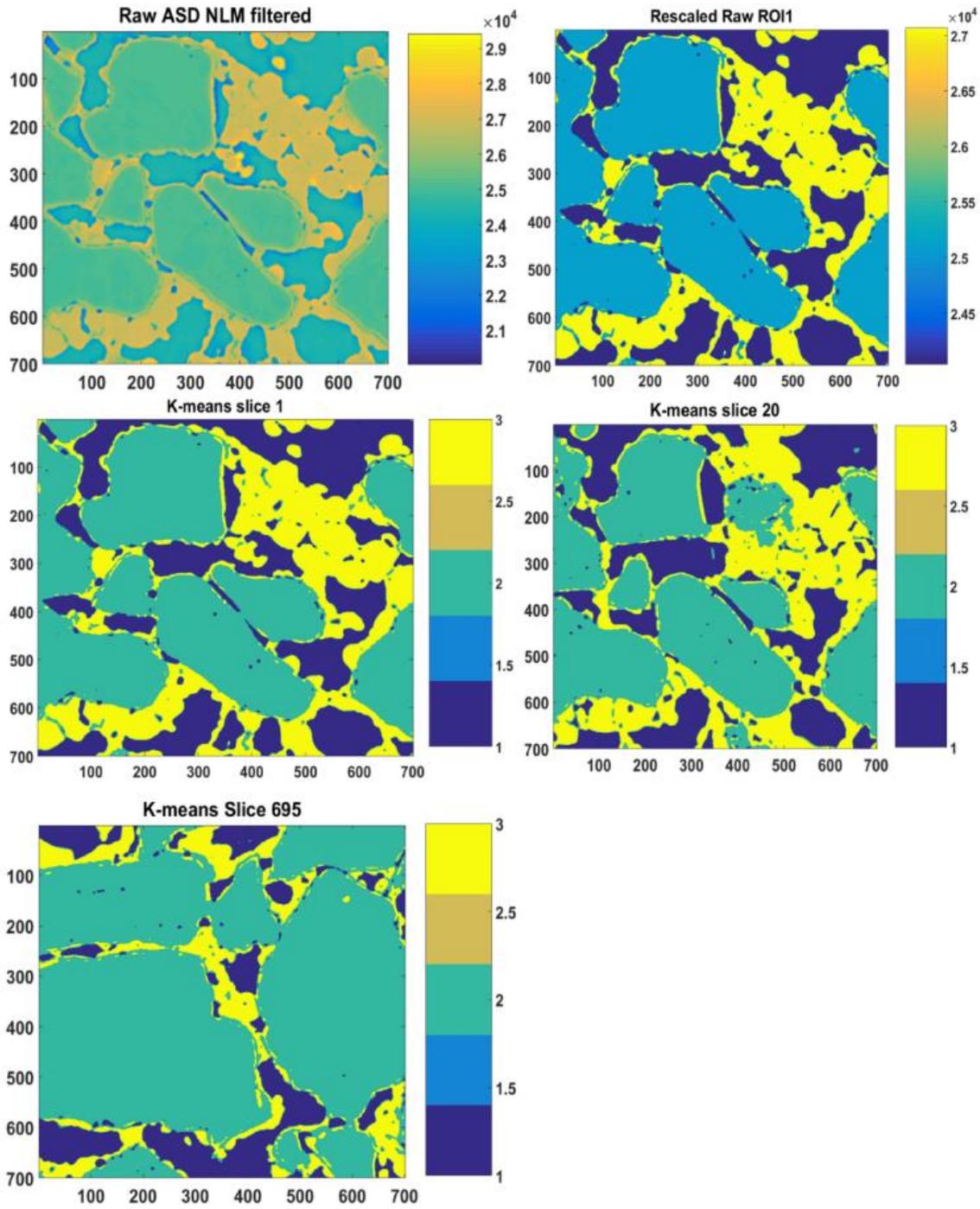

**Figure 6.** 2D filtered, rescaled, and segmented slices of gas hydrate REV1 dataset

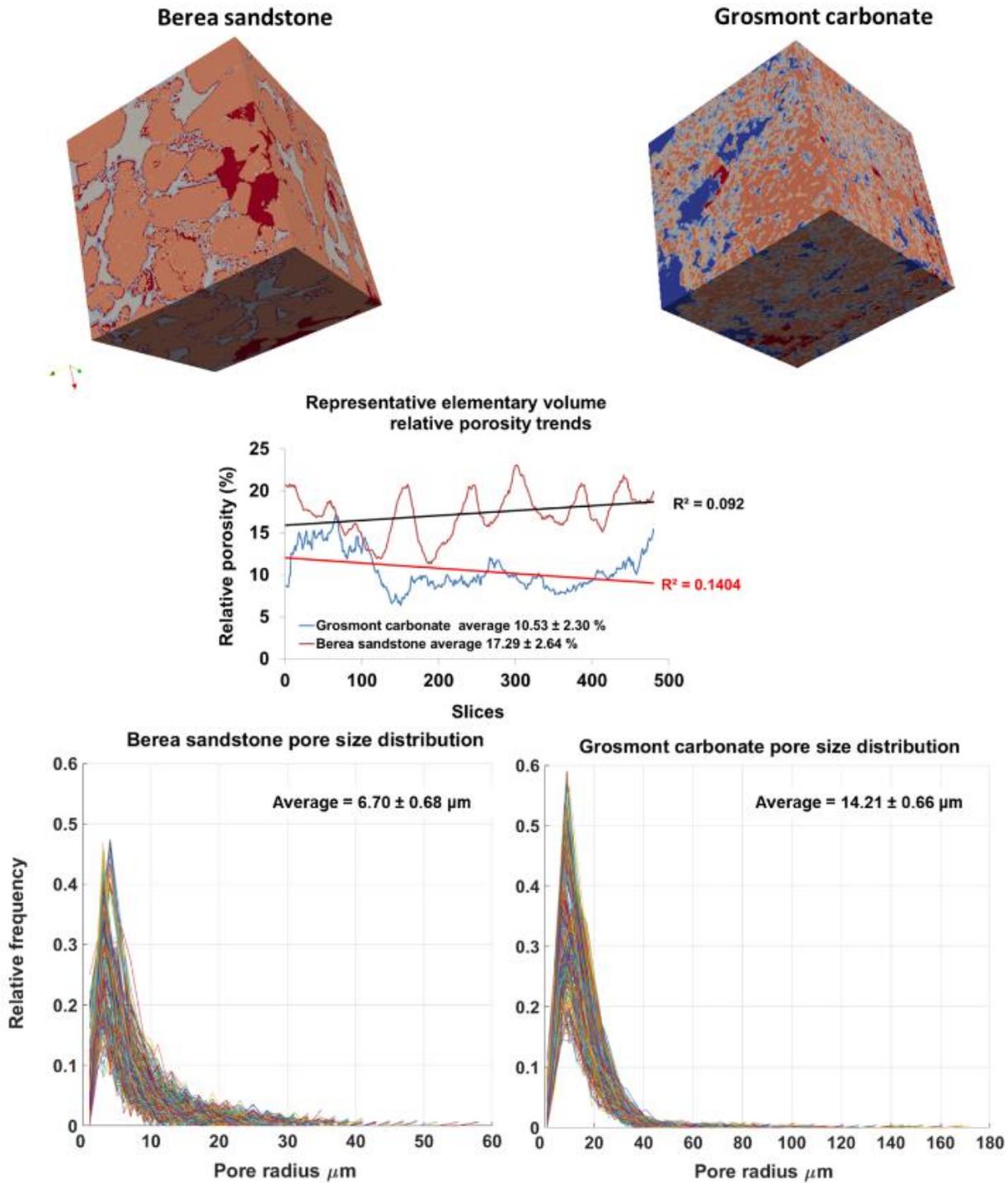

**Figure 7.** Top panel shows surface plot of REVs Berea sandstone and Grosmont carbonate (size 471x478x480) using visualisation software ParaView. Middle plot shows the relative porosity (%) trend for Berea sandstone and Grosmont carbonate REVs samples. Bottom plot shows the pore size distribution of Berea sandstone and Grosmont carbonate. XCT images were segmented using K-means. In the case of Grosmont, a non-local means filter was used

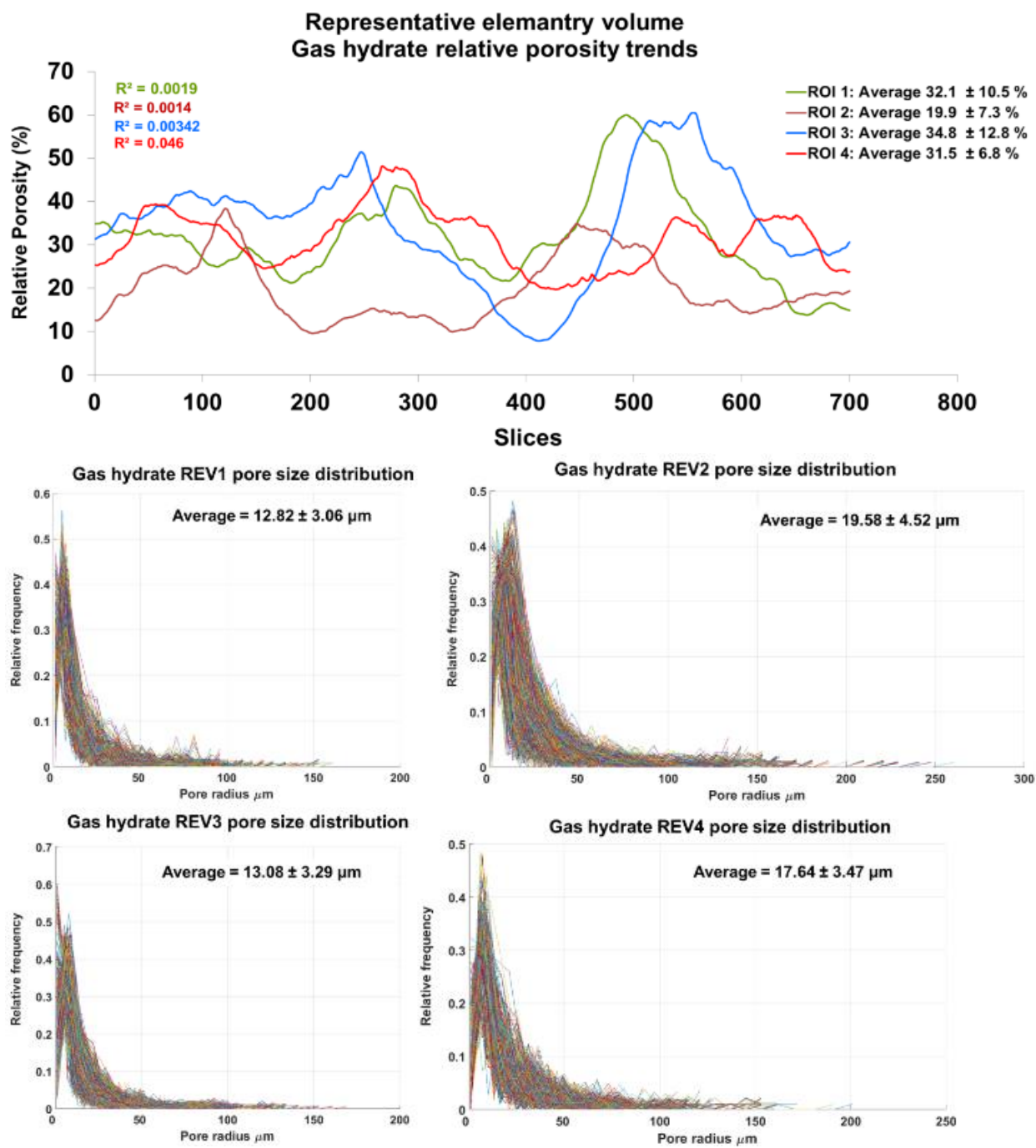

**Figure 8.** The top panel shows relative porosity trend analysis of gas hydrates, the middle and bottom panel show the geometrical pore size distribution of the respective REVs. The analysis was performed using CobWeb 1.0.

## 4.4 Estimation of Relative Porosity and Pore Size Distribution

In the case of the Grosmont carbonate and Berea sandstone, the respective REVs were segmented using K-means and LSSVM, and the PSD was calculated based on the morphological scheme suggested in Rabbani et al. (36). The mean relative porosity value of Berea sandstone is 17.3 ± 2.6 %, whereas for Grosmont carbonates the mean porosity value is lower (10.5 ± 2.3 %), as shown in Figure 7. The regression coefficient value of $R^2 = 0.092$ for the Berea sandstone porosity trend indicates that porosity remains constant throughout the REV sizes chosen and is therefore consolidated for scale-independent heterogeneities. In the case of the Grosmont carbonate rock, the chosen REV size was the best found out of five others explored, which consolidate again for scale-independent heterogeneities. The average pore size distribution thus obtained was 6.70 μm ± 0.68 μm and 14.21 μm ± 0.66 μm for Berea and Grosmont plug samples, respectively.

Similarly, the porosity and PSD of the four GH REVs were analyzed using CobWeb 1.0 except for segmentation, which was performed using a different workflow as discussed above. Figure 8 shows the comparison of the porosity trends of different GH REVs. The selected REVs consolidate for the scale independent heterogeneities. However, there is high variance compared with the mean PSD values. The exact reason is unknown, but it may be due to the drastic increase and decrease of the quartz grains, as can be noticed in Figure 6. The first and last 2D slices of ROI 1 in show either non-isotropic or isotropic distribution of quartz grains, which might have contributed to the respective high and low standard deviation seen in the porosity distribution. Figure 9 shows the surface and volume rendered plots of REV 1 and REV 2. Due to the high accuracy of segmentation, the quartz grain, brine and GH boundaries are clearly segregated and the ED effect is completely eliminated.

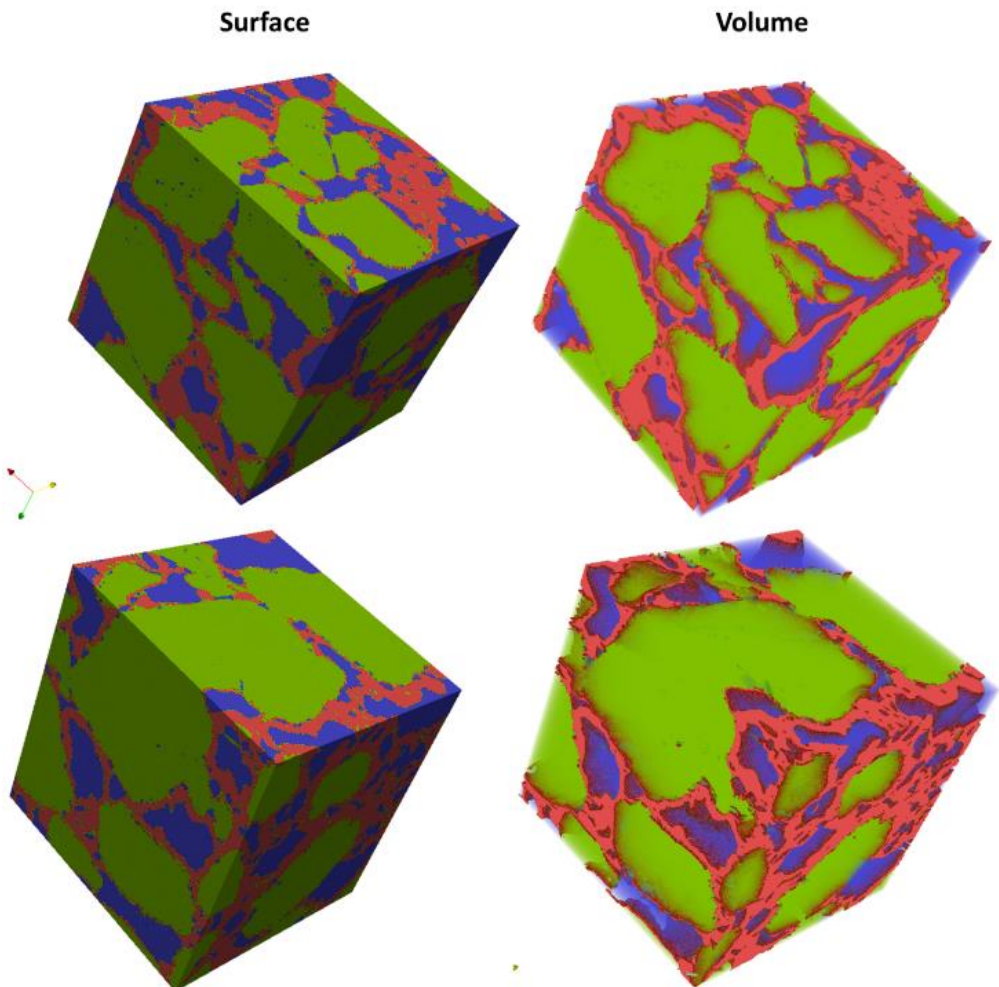

**Figure 9.** Segmented REVs of gas hydrate sample displayed as surface and volume rendered plots. as analyzed using CobWeb 1.0 and exported to VTK format using CobWeb 1.0 ParaView plug-in. Quartz grain phase is represented in green color, gas hydrate in red, and in blue the liquid brine phase.

## 5 Conclusions and Outlook

With CobWeb 1.0, this paper introduces a visualisation and image analysis toolkit dedicated to representative elementary volume analysis of digital rocks. It was developed on the MATLAB® framework and can be used as a MATLAB® plugin or as a standalone executable. It relies on robust image segmentation schemes based on machine learning (ML) techniques, which can be tested and cross-validated in parallel. Dedicated image preprocessing filters such as the non-local means, anisotropic diffusion, averaging and the contrast enhancement functions help to reduce artefacts and increase the signal to noise ratio. Petrophysical and geometrical properties such as porosity, pore size distribution and volume fractions can be computed with ease on a single representative 2D slice of a complete REV 3D stack. This was tested further using synchrotron datasets of the Berea Sandstone, a gas hydrate-bearing sediment and a tomography dataset of the Grosmont carbonate rock. The gas hydrate dataset, despite it's nanoscale resolution, was infested with strong edge enhancement (ED) artefacts, which causes discrepancies in different modelling approaches. A combination of dual filtering and dual clustering approaches is proposed to completely eliminate the ED effect in the gas hydrate sediments and the code is attached as an appendix. The REV studies performed on Berea sandstone, Grosmont carbonate rock and GH sediment using CobWeb1.0 shows relative porosity trends with very low linear regression values of 0.092, 0.1404, 0.0527 respectively. CobWeb 1.0's ability to accurately segment data without compromising data quality at a reasonable speed makes it a favourable tool for REV analysis.

However, CobWeb 1.0 is somewhat limited regarding its volume rendering capabilities, which will be one of the features to improve with the next version. The volume rendering algorithms implemented thus far are not as sophisticated as those in ParaView or DSI studio codes, which rely on the OpenGL marching cube scheme. At present, the densely nested loop structure appears to be the best choice for systematic processing. However, in future versions vectorisation and indexing approaches (*bsxfun, repmat*) will have to be tried and a check for significant changes in processing speed. MATLAB®—Java synchronisation will be explored further to configure issues related to multi-threading and visualisation (Java OpenGL).

In the science segments, the file readers and subroutines will be improved to analyse and overlay scanning electron microscope data with XCT data to enhance mineral identification. A module CrackNet (crack network) is planned which will explicitly tackle segmentation of cracks, fissures in geomaterials using machine learning techniques, and a mesh generation plugin (stl format) for 3D printing. Pore network extraction and skeletonisation schemes such as modified maximum ball algorithm (37) and medial axis transformation (38) will be considered such that the data can be exported to an open-source pore network modelling package (39).

## 6 Acknowledgements

We thank Heiko Andrä and his team at Fraunhofer ITWM, Kaiserslautern, Germany, for providing us with the synchrotron tomography dataset of the Berea sandstone. The acquisition of the gad hydrate synchrotron-data was funded by the German Science Foundation (DFG grant Ke 508/20 and Ku 920/18). This study was funded within the framework of the SUGAR (Submarine Gashydrat Ressourcen) III project by the Germany Federal Ministry of Education and Research (BMBF grant 03SX38IH). The sole responsibility of the paper lies with the authors.

## 8 Appendix A: MATLAB snippet for removal for Edge Enhancement Effect in gas hydrate datasets

### 8.1 Gas Hydrate Segmentation

### 8.2 Step 1

The Dual Clustering approach, by which first the 16bit gas hydrate was filtered using Anisotropic diffusion (ASD) filter, and then with non-local means (NLM) filter, to minimize/normalize the edge enhancement (ED) artefacts.

### 8.3 Step 2

- read slice by slice 3D prefiltered raw data
- for this example the reading is restricted
- to only four slices (700x700x4); it can be changed using nZ variable

```matlab
mfname= 'Xe_17w_8_ROI4_ADS_NLM';
ifname=[mfname,'.raw'];
nX=700;
nY=700;
nZ=4;
ldim = 1;
xDi=[nX nY]';
grenzwert=0;
clusterS =7;
ifid=fopen(ifname,'r');
M=zeros(nX,nY,nZ,'uint16');
SeData = nZ-ldim;
SeData  = 1:1:SeData;
dim = size(M);

for k=ldim:nZ
    disp(sprintf('Reading slice no. %d....',k));
    s=sprintf('Slice: % d', k');
    S=fread(ifid, [xDi(1) xDi(2)], 'uint16');
    M(:,:,k)=S;
    %figure; imagesc(M(:,:,k)); colorbar;
end
```

```
Reading slice no. 1....
Reading slice no. 2....
Reading slice no. 3....
Reading slice no. 4....
```

### 8.4 Display image

```matlab
figure; imagesc(M(:,:,1)); colorbar;
title('Raw Prefiltered ASD NLM')
```

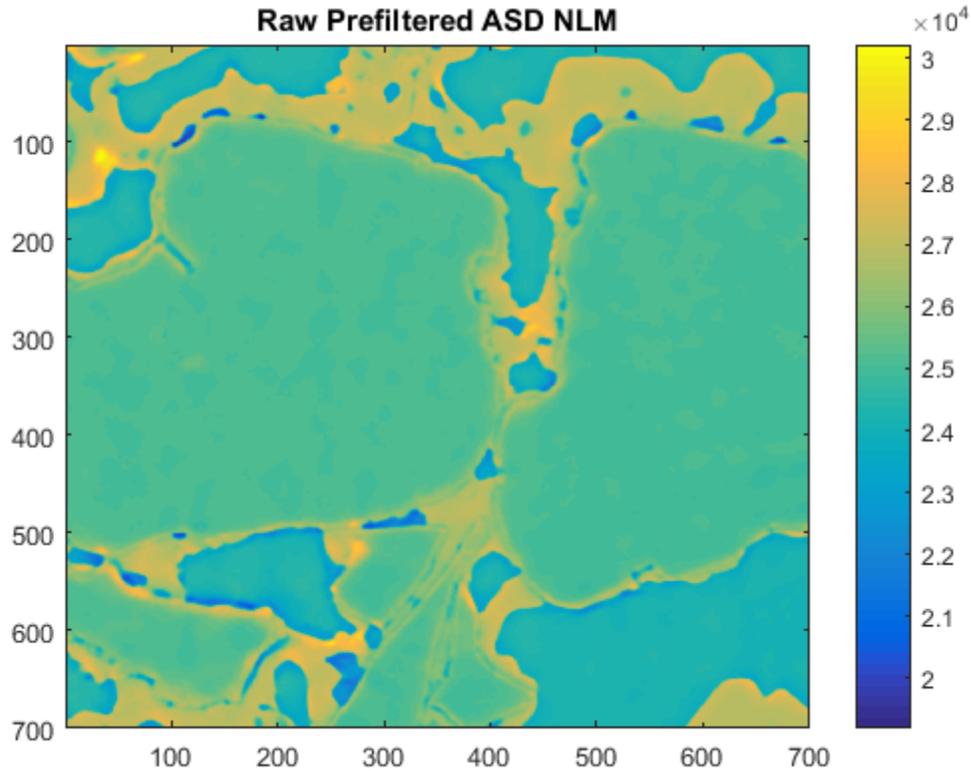

### 8.5 Concatenate raw data into single array

```
%* concatenate array will be used in step three

M = M(:,:,ldim:nZ);
rawM = double(M(:));
```

### 8.6 Perform k-means clustering

Here, clustering is restricted to class 7 optimal to enable clustering of all the available features:

```
for ii = 1:2
    R=double(M(:,:,ii));
    [r,c,v]=find(R>grenzwert);
    cyl=R>grenzwert;
    R1=cyl.*R;
    [m, n, w]=find(R1);
    G = kmeansK(w,clusterS);
    S=sparse(r,c,G,size(R,1),size(R,2));
    M_seg=full(S);
    SegImg(:,:,ii)=M_seg;
    %figure; imagesc(SegImg(:,:,ii)); colormap(parula(5)); colorbar;
    %title('K-means prefiltered');
end
```

21## 8.7 Display image

```
figure; h = imagesc(SegImg(:,:,1)); colormap(jet(max(h.CData(:))));
title('K-means prefiltred');
```

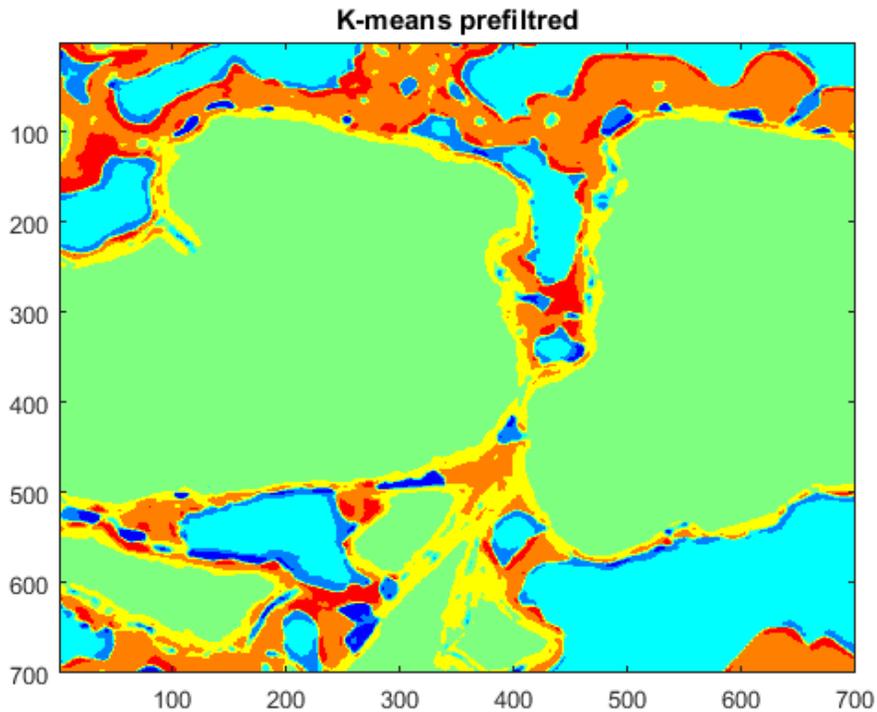

## 8.8 Step three

The purpose is to |index out| pixel values of different phases:

```
% noise
% edge enhanced low (EDL)
% liquid
% quartz
% edge enhanced high (EDL)
% gas hydrate
% from the concatenated raw images matrix using segmented class values
% thereafter compare their histogram % _as sanity check_
% to identify if any overlapping boundaries
```

## 8.9 Index noise pixels

```
rangeNl = 0;
indN = find(h.CData(:)==rangeNl);
rawO = rawM(indN);
```

## 8.10 Plot histogram noise

```
[cN, countN] = hist(rawO, 10);
%figure; bar(countN, cN);
%title('noise')
```

## 8.11 Index EDL pixels

```
rangeNu = 2;
indD = find(h.CData(:)>rangeNl & h.CData(:)<=rangeNu);
rawD = rawM(indD);
```

## 8.12 Plot histogram noise

```
[cD, countD] = hist(rawD, 100);
%figure; bar(countD, cD);
%title('Edge Enhanced low noise')
```

## 8.13 Index liquid pixels

```
rangeLl = 1;
rangeLu = 3;

indL = find(h.CData(:)>=rangeLl & h.CData(:)<=rangeLu);
if min(SegImg(indL))==rangeLl & max(SegImg(indL))==rangeLu
    rawL = rawM(indL);
    min_rawL = min(rawL);
    max_rawL = max(rawL);
    Avg_rawL = mean(rawL);
else
    fprintf('min and max for liquid dont match.....\n')
    return
end
```

## 8.14 Plot histogram liquid

```
[cL, countL] = hist(rawL, 100);
figure; bar(countL, cL);
title('Liquid')
```

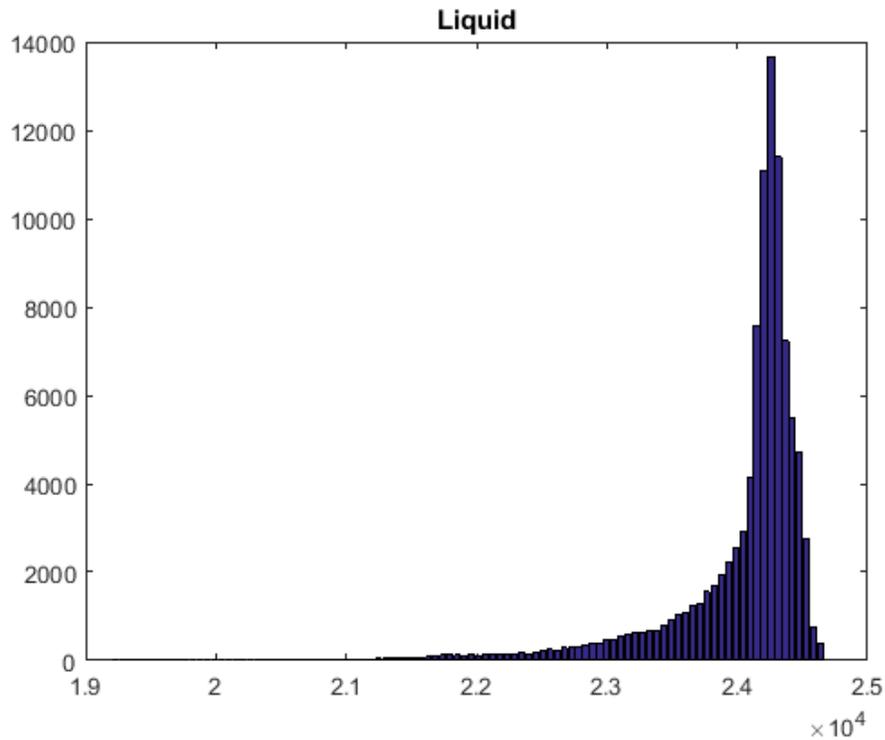

## 8.15 Index EDH pixels

```
rangeE = 5;
indE = find(h.CData(:)==rangeE);
if min(SegImg(indE))==rangeE
    rawE = rawM(indE);
    min_rawE = min(rawE);
    max_rawE = max(rawE);
    Avg_rawE = mean(rawE);
else
    fprintf('min and max for EDH dont match.....\n')
    return
end
```

## 8.16 Plot histogram EDH

```
[cE, countE] = hist(rawE, 10);
figure; bar(countE, cE);
title('Edge Enhanced high noise')
```

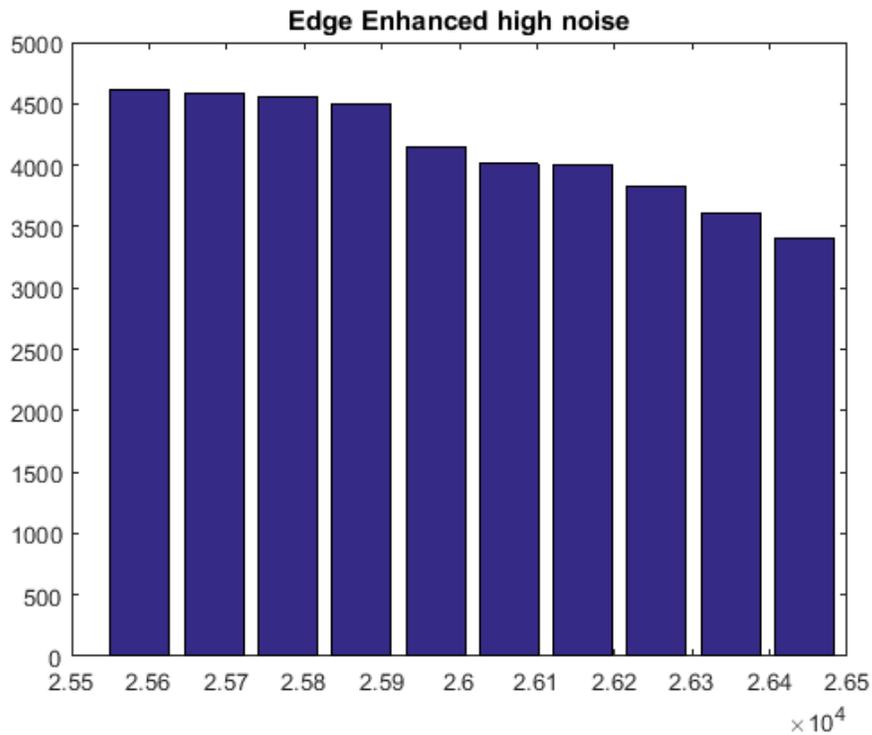

## 8.17 Quartz index phases

```
rangeQu = 4;
```

## 8.18 Quartz

```
indQ = find(h.CData(:)==rangeQu);
if min(SegImg(indQ)) == rangeQu
    rawQ = rawM(indQ);
    min_rawQ = min(rawQ);
    max_rawQ = max(rawQ);
    Avg_rawQ = mean(rawQ);
else
    fprintf('min and max for quartz dont match.....\n')
    return
end
%indQ = find(h.CData(:)>=rangeQl & h.CData(:)<=rangeQu);
```

## 8.19 Plot histogram quartz

```
[cQ, countQ] = hist(rawQ, 100);
figure; bar(countQ, cQ);
title('Quartz')
```

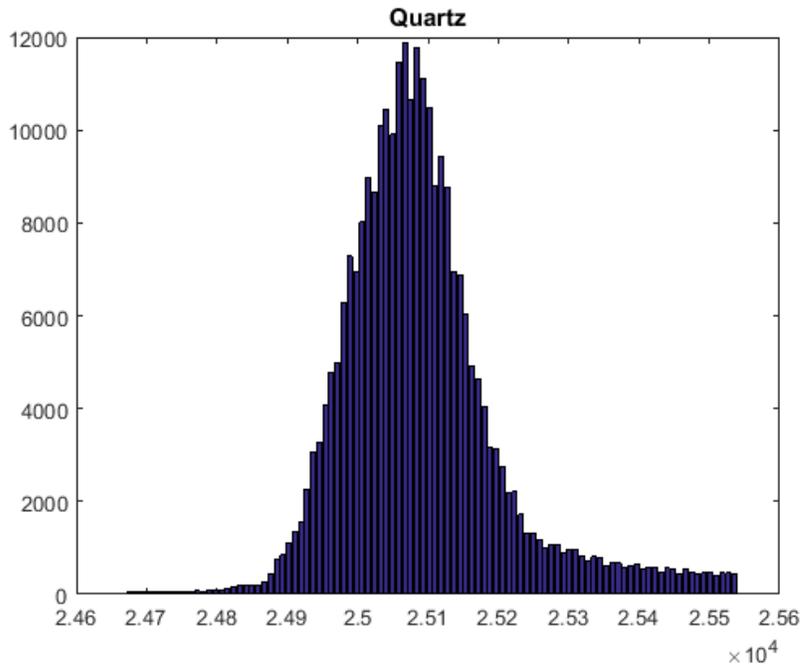

## 8.20 Gas Hydrate

```
rangeMl =6;
rangeMu =7;
%indM =find(h.CData(:)>=rangeMu);
indM = find(h.CData(:)>=rangeMl & h.CData(:)<=rangeMu);
if min(SegImg(indM))==rangeMl & max(SegImg(indM))==rangeMu
    rawMu = rawM(indM);
    min_rawMu = min(rawMu);
    max_rawMu = max(rawMu);
    Avg_rawMu = mean(rawMu);
elseif min(SegImg(indM))==rangeMu & max(SegImg(indM))==rangeMu
    rawMu = rawM(indM);
    min_rawMu = min(rawMu);
    max_rawMu = max(rawMu);
    Avg_rawMu = mean(rawMu);
else
    fprintf('min and max for gas hydrate dont match.....\n')
    return
end
```

## 8.21 Plot Histogram Gas Hydrate

```
[cM, countM] = hist(rawMu, 100);
figure; bar(countM, cM);
title('Methane')
```

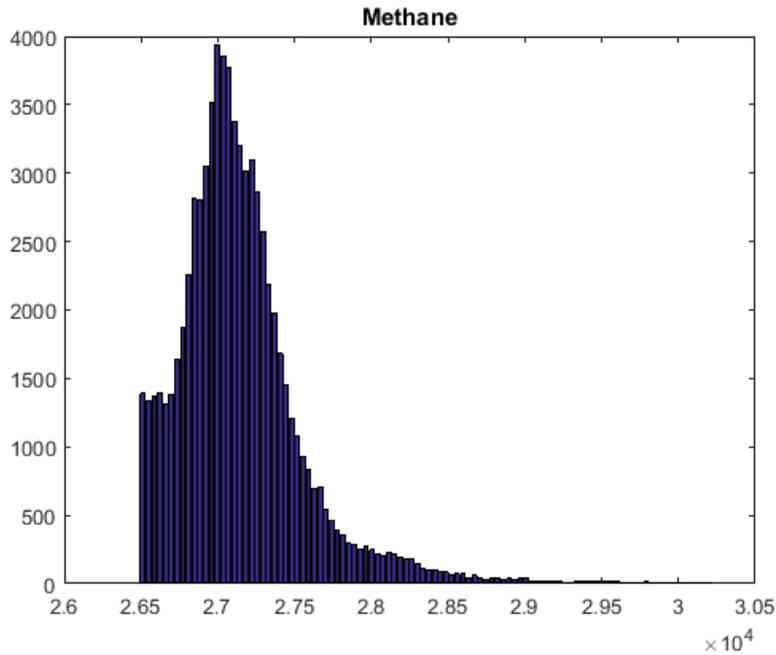

## 8.22 Step 4 - rescaling the raw images

First min-max and mean of respective phases are calculated for the respective (raw) phases (obtained above), which thereafter are replaced by their respective mean values.

```
%average values
```

## 8.23 With an exception to GH dataset

Where EDH (raw pixels) are replaced with averaged quartz values:

```
%as they are in close vicinity to quartz pixel values

M_replace = M(:);
min_li = min_rawL;
max_li =max_rawL;
avg_li =Avg_rawL;
min_Qz = min_rawQ;
max_Qz = max_rawQ;
avg_Qz = Avg_rawQ;
min_EDH = min_rawE;
max_EDH = max_rawE;
avg_EDH = Avg_rawE;
min_GH = min_rawMu;
max_GH = max_rawMu;
avg_GH = Avg_rawMu;

%-----------------------------------------------------------------------
% indxes of liquid pixels
%-----------------------------------------------------------------------
Ind_rep_L = find(M_replace>=min_li & M_replace <= max_li);
% replacement by average liquid value
if min(M_replace(Ind_rep_L))==min_li & max(M_replace(Ind_rep_L)==max_li)
    M_replace(Ind_rep_L)=avg_li;
else
    fprintf('min and max for liquid dont match.....\n')
    return
end

%-----------------------------------------------------------------------
% indxes of quartz pixels
%-----------------------------------------------------------------------
```

```matlab
Ind_rep_Q = find(M_replace>= min_Qz & M_replace<= max_Qz);
% replacement by average quartz value
if min(M_replace(Ind_rep_Q))==min_Qz & max(M_replace(Ind_rep_Q))==max_Qz
    M_replace(Ind_rep_Q)= avg_Qz;
else
    fprintf('min and max for quartz dont match.....\n')
    return
end
%-------------------------------------------------------------------------
% indxes of EDH pxels
%-------------------------------------------------------------------------
Ind_rep_E = find(M_replace>= min_EDH & M_replace<=max_EDH);
%
```

### Replace by average quartz values

```matlab
if min(M_replace(Ind_rep_E))==min_EDH & max(M_replace(Ind_rep_E))==max_EDH
    M_replace(Ind_rep_E)= avg_Qz;
else
    fprintf('min and max for EDH dont match.....\n')
    return
end

%-------------------------------------------------------------------------
%indexes of gas hydrate pixels
%-------------------------------------------------------------------------
Ind_rep_M = find(M_replace>=min_GH & M_replace<=max_GH);
% replacement by average gas hydrate value
if min(M_replace(Ind_rep_M))== min_GH & max(M_replace(Ind_rep_M))==max_GH
    M_replace(Ind_rep_M)= avg_GH;
else
    fprintf('min and max for methane dont match.....\n')
    return
end

%-------------------------------------------------------------------------
```

## 8.24 Reshape rescaled array

```matlab
%-------------------------------------------------------------------------
M_replaced = reshape(M_replace,[dim(1), dim(2), dim(3)]);
clear M_replace;
figure; imagesc(M_replaced(:,:,1));
title('Rescaled Raw');
```

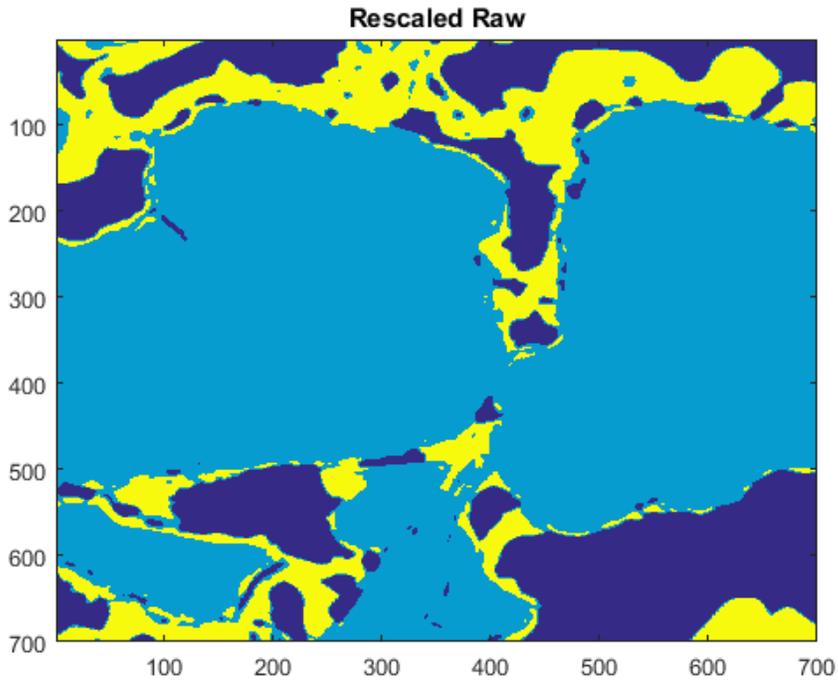

### 8.25 Step 5

K-means clustering is performed on the rescaled images to obtain segmetation in three classes:

```
clusterS =3;
initialcenters = [avg_li,avg_Qz,avg_GH];
for ii = 1:dim(3)
    R=double(M_replaced(:,:,ii));
    [r,c,v]=find(R>grenzwert);
    cyl=R>grenzwert;
    R1=cyl.*R;
    [m, n, w]=find(R1);
    G = kmeans(w,clusterS,'Distance','sqeuclidean','start',initialcenters');
    S=sparse(r,c,G,size(R,1),size(R,2));
    M_seg=full(S);
    SegImg(:,:,ii)=M_seg;
    %figure; imagesc(SegImg(:,:,ii)); colormap(parula(5)); colorbar;
end

figure; imagesc(SegImg(:,:,1)); colormap(parula(5)); colorbar;
title('Gas Hydrate ROI');

fclose('all');
```

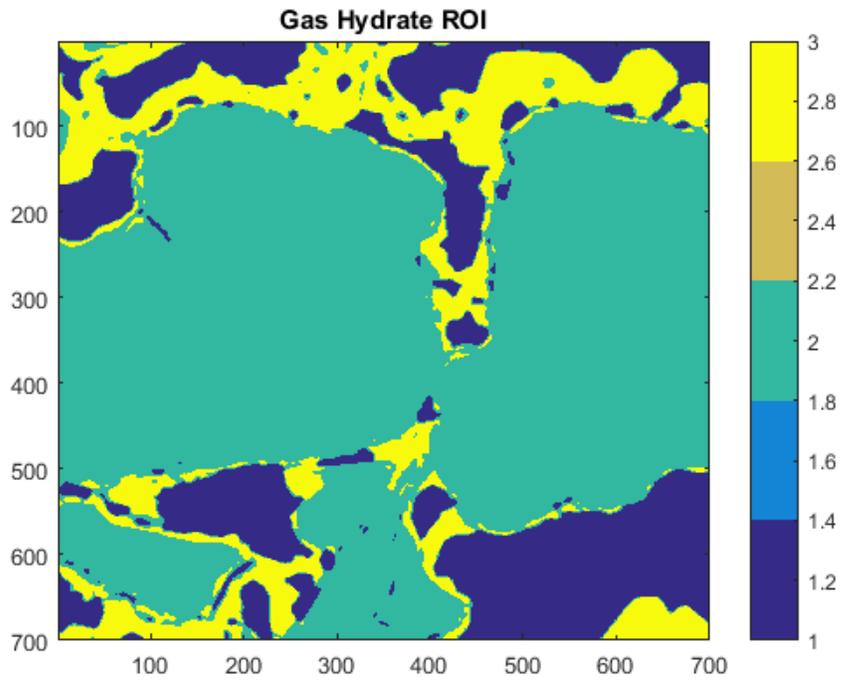

*Published with MATLAB® R2015b*